\documentstyle[12pt]{article}
\begin{document}
\begin{titlepage}
\begin{flushright}
CIEA-GR-9401
\end{flushright}
\begin{center}
\vskip 3em
{\Large Super-Minisuperspace and New Variables}
\vskip 2em
{\large Riccardo Capovilla ${}^{(1)}$ and Jemal Guven ${}^{(2)}$}\\[1em]
\em{${}^{(1)}$ Departamento de F\'isica \\
Centro de Investigaci\'on y de Estudios Avanzados, I.P.N.\\
Apdo. Postal 14-740, Mexico 14, D.F., Mexico.\\[.7em]
${}^{(2)}$ Instituto de Ciencias Nucleares\\
Universidad Nacional Autonoma de Mexico \\
Apdo. Postal 70-543 Circuito Exterior, C.U., Mexico D.F., Mexico.}

\end{center}
\vskip 1em
\begin{abstract}
We consider the specialization
to spatially homogenous solutions of the
 Jacobson formulation of N=1 canonical
 supergravity in terms of Ashtekar's new variables.
 We find that the classical Poisson algebra of
 the supersymmetry constraints
is preserved by this specialization
only for Bianchi type A models.
The quantization of supersymmetric Bianchi type A models
is carried out in the triad representation. We find the
physical states of this quantum
theory. Since we are missing a suitable inner product on these
physical states, our results are only formal.

\vspace{1cm}


\end{abstract}
\end{titlepage}
\newpage


\def\eg{{\it e.g.${~}$}}
\def\ie{{\it i.e.${~}$}}
\def\frac#1#2{{\textstyle{#1\over\vphantom2\smash{\raise.20ex
        \hbox{$\scriptstyle{#2}$}}}}}
\def\st{\widetilde{\sigma}}
\def\pt{\widetilde{\pi}}
\def\e{\varepsilon}
\def\D{{\cal{D}}}
\def\p{\psi}
\def\s{\sigma}
\def\H{{\cal{H}}}
\def\J{{\cal{J}}}
\def\Sc{{\cal{S}}}
\def\Sh{{\cal{S}}^{\dagger A}}
\def\ph{\psi^{\dagger}}


\def\ga{\mbox{\bf {\scriptsize{a}}}}
\def\gb{\mbox{\bf {\scriptsize{b}}}}
\def\gc{\mbox{\bf {\scriptsize{c}}}}
\def\gd{\mbox{\bf {\scriptsize{d}}}}
\def\ge{\mbox{\bf {\scriptsize{e}}}}
\def\gf{\mbox{\bf {\scriptsize{f}}}}
\def\gg{\mbox{\bf {\scriptsize{g}}}}
\def\gh{\mbox{\bf {\scriptsize{h}}}}
\def\gi{\mbox{\bf {\scriptsize{i}}}}
\def\gj{\mbox{\bf {\scriptsize{j}}}}


\noindent{\bf 1. Introduction}

\vspace{.3cm}

Witten's proof of the positivity of energy
in general relativity is a strong indication that
supergravity has something interesting to say about
general relativity \cite{Witten,Deser}.
In the same spirit, supergravity can be put to use in quantum
cosmology. An interesting application is the investigation of the
physical states of the quantum theory of minisuperspace models.
In supergravity, the generators of
supersymmetry transformations may be seen as `square roots'
of the generators of diffeomorphisms that make up the
hamiltonian for vacuum general relativity \cite{Teitelboim}. Thus,
modulo factor ordering ambiguities,
physical states for supergravity, \ie states annihilated by
the supersymmetry generators, are also
physical states for general relativity,
although the converse
is of course not true.  By appealing to
supergravity, one would like to identify the class of
priviledged  physical states that are  supersymmetry invariant.

In the context of quantum cosmology, using canonical quantization
methods, this line of research
has been pursued using two different approaches.
The first considers the  Wheeler-De Witt equation for some
vacuum Bianchi model, and then takes its `square root', relying
on the analogy between the Wheeler-De Witt equation and the equation
of motion for a relativistic particle (see \eg \cite{Gra}).
This approach can be implemented only within a specific Bianchi
model, and it has the disadvantage of introducing fermionic variables
without a clear physical interpretation. Recently,
it has been exploited in the
Ashtekar formulation of canonical gravity in \cite{OPR}.

A second approach, which we will follow, is to consider
the specialization of supergravity to the case of spatially
homogenous solutions. We call this specialization super-minisuperspace.
It is the generalization to supergravity of Misner's minisuperspace
\cite{Misner}.
This approach has the benefit of providing a straightforward
identification of the phase
space variables, and of their transformations properties.

Using the triad extended ADM canonical formalism for supergravity \cite{DEC},
various  minisuperspace models have been studied, \eg Bianchi I in
\cite{MOR,DEHO}, Taub in \cite{SOM},
diagonal Bianchi IX in \cite{MOS,DEIX}. The results of these investigations
indicate
 that the quantum supersymmetry constraints select
 only the most symmetrical quantum states.
(In \cite{DEF}, D'Eath comes to the same conclusion for physical states
with a finite number of fermionic fields for {\it full} supergravity.
However, both his result and its interpretation are still
controversial \cite{Page}.)

In this paper, we consider the super-minisuperspace of
hamiltonian N=1 supergravity in the formulation
 given by Jacobson \cite{Jac}.
This formulation extends to supergravity the Ashtekar
formulation for canonical gravity \cite{Ash}. The spatial left-handed
spin connection, and the densitized spatial triad are used as bosonic phase
space variables.
The Jacobson formulation shares many of the
nice features of Ashtekar's formulation of gravity. In
particular, one is using conjugate phase space variables, and
the constraints are put in polynomial form
of low order. As a result, the canonical quantization
is much
easier to carry  out than in the  triad extended ADM
formalism, where one is working with awkword non-conjugate
phase space variables. Exploiting these simplifications, we carry out
the  canonical quantization of
super-minisuperspace for {\it all} type A Bianchi models.
Moreover, we consider the most general factor
ordering of the quantum constraints \footnote{After this work was
completed we became aware of Ref. \cite{ATY}, where the
quantization of type A Bainchi models is carried out in
the triad ADM formalism.}.

A key technical
assumption in our analysis, and in previous treatments,
is that quantum states
depend only on a {\it finite}
number of fermionic fields. This permits one to expand the
wavefunction in a finite number of terms corresponding
to even powers of the fermionic fields, since they are
grassmannian. This assumption is not reasonable
if one is interested in the full space of physical
states for supergravity, where states with an infinite
number of fermionic fields are likely to play a key
role \cite{DEF,Jacpriv}. However, the assumption is
justified in the super-minisuperspace approximation, where
one is dealing only with a finite number of degrees
of freedom, and, in any case, if one is interested only in the
limit in which the fermionic fields vanish, \ie general
relativity.

Various representations of the quantized theory are possible.
In this paper, we focus on the triad representation.
Quantum states are represented by wavefunctions
that depend on the spatial triad, and on the spatial gravitino
field.  Our reason for priviledging this representation is
that it is the closest to the representation used in previous treatments
which employed the
triad  ADM formalism.

We find that physical states for quantum Bianchi type A models are in general
 of the following form.
The bosonic part of the wavefunction annihilated by the quantum
consdtraints is given by
\begin{displaymath}
\Psi_{(0)} = k_{(0)} h^{\alpha / 2} exp \{ 2i  F\},
 \end{displaymath}
where $k_{(0)}$ is a constant, $h_{\ga \gb}$ is the
homogenous metric, $h:= det h_{\ga \gb}$. The constant $\alpha$
is determined by the factor ordering chosen. $F$ is the bosonic
part of the homogenous specialization of the generating function of the
canonical transformation from the ADM variables to Jacobson's.
($F$ is pure
imaginary.)
In general, the next terms in the wavefunction, $\Psi_{(2)}$ and
$\Psi_{(4)}$, quadratic and quartic in the fermionic fields, respectively,
are found to vanish. The exceptions require a fine adjustement of the
factor ordering.
The last term, of sixth order in the fermionic fields,
is given by
\begin{displaymath}
\Psi_{(6)} = k_{(6)} h^{1/3} [\beta^2] [\rho^2 ]^2,
\end{displaymath}
where $k_{(6)} $ is a constant, and $ [\beta^2], [\rho^2]$ are Lorentz
invariants built with the gravitino field, defined in the text below.
This term of the wavefunction
has the interesting feature
of being independent of the specific Bianchi model, and of the factor
ordering.

We emphasize that our resuls are  only formal.
Although we identify the  space of physical states, we have
not identified an inner product on this space. Therefore,
we are unable to extract any physical information
from these quantum states.

The use of the left-handed connection  as phase space variable,
implies the use of complex
variables.  To recover the real phase space, one needs to impose  reality
conditions.
In turn, these reality conditions can be used to select an inner product
on the physical states \cite{Ashbook,Tate}. Since the system
is finite-dimensional, it should be possible to carry out this
step \cite{Rendall}, but we have not tried to do it here, leaving
this problem for future work.

We also leave for future work quantization in  the connection representation.
For this representation, the supersymmetry constraints are
second order in the momenta, as for pure gravity. Thus,
much of the motivation for looking at supergravity is
lost. However, it may still be  of interest as a first
timid step towards the inclusion of matter fields
in the new variables non-perturbative quantization of
gravity.

This paper is organized as follows. In sect. 2, we review briefly the
Jacobson canonical formulation of N=1 supergravity. In sect. 3, we
summarize the kinematics of spatial homogeneity. The specialization to
homogenous supergravity is decribed in sect. 4. In sect. 5, we
consider the canonical quantization of Bianchi type A models in the
triad respresentation, and we solve explicitly the quantum
supersymmetry constraints.

\vspace{.5cm}

\noindent{\bf 2. Jacobson's Hamiltonian N=1 Supergravity}

\vspace{.3cm}

In this section, we briefly recall the
Jacobson formulation of hamiltonian
N=1 supergravity in terms of Ashtekar's new
variables \cite{Jac}.
We refer the reader to \cite{Jac} for its derivation from an action
principle, and for
a thorough discussion.

The canonical coordinates are a  complex traceless
$SL(2,C)$ spatial connection,
$A_{iA}{}^B$, and a traceless spatial vector density of weight 1,
$\st^{iAB}$, together with the spatial
anti-commuting gravitino field $\p_i^A$, and its conjugate momentum
the anti-commuting
$\pt^{iA}$. (Small latin letters from the middle of the alphabet
denote spatial indices, $i,j,\dots = 1,2,3$. Capital latin letters
denote $SL(2,C)$ indices $A,B,\dots = 0,1$. These indices are raised
and lowered with the anti-symmetric symbol $\epsilon^{AB}$, and its
inverse $\epsilon_{AB}$, according to the rules $\lambda^A =
\epsilon^{AB} \lambda_B$, $\lambda_A = \lambda^B \epsilon_{BA}$.)

The connection $A_{iA}{}^B$ is the spatial pull-back of the
left-handed spin connection. The vector density
$\st^{iAB}$ may be interpreted as the (densitized) spatial
triad, in the sense that the covariant
(doubly densitized) spatial metric is given by
$ (det q) q^{ij} = \st^{i AB} \st^j{}_{AB} $.
The momentum $\pt^i_A $ is related to the complex conjugate of the
spatial gravitino field, $\psi_i^A$.

The fundamental Poisson brackets are given by
\begin{eqnarray}
\{ \st^{kAB}(x), A_{jCD} (y) \} &=&  {i \over \sqrt{2}}
\delta_j^k \delta_{(C}{}^A
\delta_{D)}{}^B \delta^3(x,y) , \label{eq:pb1} \\
\{ \pt^{kA}(x), \p_{jB} (y) \} &=&  {i \over \sqrt{2}} \delta_j^k \delta_B{}^A
 \delta^3(x,y). \label{eq:pb2}
\end{eqnarray}
Note that we differ with Jacobson by a factor of $i/\sqrt{2}$ on the right
hand side.

In terms of these variables, the hamiltonian for N=1
(complex) supergravity may be written in
{\it polynomial} form, at most quartic in the phase space variables,
\begin{equation}
H = i \sqrt{2} \int d^3 x  \{ e_{0AB} \H^{AB} + \p_{0A} \Sc^A
+ \Sh \ph_{0A} + A_{0AB} \J^{AB} \}.
\label{eq:ham}
\end{equation}
The fields $ e_{0AB}, p_{0A}, \ph_{0A} , A_{0AB} $
are Lagrange multipliers, that enforce the
constraints
\begin{eqnarray}
\H^{AB} :&=& (\st^i \st^j F_{ij} )^{ BA} +
2 (\pt^j \st^k \D_{[j} \p_{k]} )\e^{AB}
+ 2 (\pt^j \D_{[j} \p_{k]}) \st^{kAB} = 0,
\label{eq:diffeo}\\
\J^{AB} :&=& \D_k \st^{kAB} - \pt^{k(A} \psi_k^{B)} = 0  ,
\label{eq:gauss} \\
\Sc^A :&=& \D_k \pt^{kA} = 0 ,
\label{eq:susy1} \\
\Sh :&=& (\st^j \st^k \D_{[j} \p_{k]} )^A = 0 .
\label{eq:susy2}
\end{eqnarray}
Here $\D_i$ is the covariant derivative of $A_{iA}{}^B$, with curvature
$F_{ij A}{}^B := 2 \partial_{[i} A_{j]A}{}^B  +
2 A_{[i | A}{}^C A_{j]C}{}^B$. Following Jacobson, we are
using the convention that
suppressed spinor indices are contracted from
upper left to lower right, \eg $ ( \st^i \st^j )^{AB} = \st^{iAC} \st^j{}_C{}^B
$.
Hermitian conjugation is defined with respect to some
Hermitian metric $n^{AA'}$.

The constraint $\H^{AB} = 0 $  is the generator of
diffeomorphisms. Its part symmetric in the indices $AB$ generates
diffeomorphisms tangential to the space-like hypersurface,
up to an $SL(2,C)$ rotation, and up to a right-handed supersymmetry
transformation.
$\H^{AB} \e_{AB} $ generates diffeomorphisms out of the
hyper-surface up to a right-handed supersymmetry
transformation\footnote{The consequences of writing the
diffeomorphism  constraints in this `unified' form have
been studied in  Ref. \cite{JacRom}.}.
See \cite{Jac} for its derivation. Basically, it
is the appropriate linear combination of the
diffeomorphism constraint and a constraint of the form (\ref{eq:susy2})
that makes the hamiltonian polynomial in the phase space variables.

The constraint $\J^{AB}$ enforces $SL(2,C)$ covariance, and generalizes
the Gauss constraint of vacuum general relativity.

The last two constraints, $\Sc^A, \Sh $,
generate left-handed, and right-handed supersymmetry
transformations, respectively.
Their Poisson algebra reads,
\begin{eqnarray}
\{ \Sc^A , \Sc^B \} &=& 0 ,\label{eq:pas1}\\
\{ \Sh , \Sc^{\dagger B } \} &=&  C^{AB}{}_C \Sc^{\dagger C } ,
\label{eq:pas2}\\
\{ \Sc^A , \Sc^{\dagger B } \} &=& {i \over 2 \sqrt{2}} \H^{AB} ,
\label{eq:pas3}
\end{eqnarray}
where $ C^{AB}{}_C := (i / \sqrt{2}) (2 \st^{k (A}{}_C \psi_k^{B)} - \st^{kAB}
\psi_{kC} )$.

The interesting bracket is the last one. The
mixed Poisson bracket gives the `unified' diffeomorphism
constraint (\ref{eq:diffeo}). In this sense, one can
think of the supersymmetry constraints as a `square root'
of the diffeomorphism constraints \cite{Teitelboim}.

As in the case of pure general relativity, the use of the
self-dual spin connection as field variable implies the use
of {\it complex} coordinates on the
real phase space of supergravity.
Therefore, to recover the physical phase space,
one needs to introduce  appropriate reality
conditions. One can require $\st^{iAB}$ to be Hermitian with respect
to some Hermitian metric $n^{AA'}$.  This gives reality conditions for the
fermionic fields, that amount to the relation of $\pt^{iA}$ with the
complex conjugate of $\psi_i{}^A$.
To impose reality condition on the connection  two options are
available. The first is non-polynomial,
and may be written as,
\begin{equation}
A_i{}^{AB} = \Gamma_i{}^{AB} + i \Pi_i{}^{AB} ,
\label{eq:rc1}
\end{equation}
where $\Gamma_i{}^{AB}$ is the
Hermitian (torsionful) spin connection compatible
with $\st^{iAB}$, and  $\Pi_i{}^{AB}$ is Hermitian, and determined by
the extrinsic curvature.
The second option is to  impose  polynomial conditions equivalent to
(\ref{eq:rc1})
by requiring that the reality of the  (doubly densitized) metric  $q q^{ij}$
be preserved in time \cite{ART}.

\vspace{.5cm}

\noindent{\bf 3. Bianchi formalism}

\vspace{.3cm}

In this section, we summarize the kinematics of
spatial homogeneity. We
follow closely the analogous specialization of
the new variables formulation of hamiltonian
vacuum general relativity given by Ashtekar and Pullin
in \cite{AshtekarPullin}.

As in the ADM treatment of Bianchi models
(see \eg \cite{MacCallum,RyanShepley}), we consider a
kinematical triad of vectors, $X^i_{\ga}$, which commute
with the three Killing vectors on the
spatial hypersurface $\Sigma$. (Latin letters from the
beginning of the alphabet, $a,b,c,...$
label the triad vectors.)
The triad satisfies,
\begin{equation}
[ X_{\ga} , X_{\gb} ]^i  = C_{\ga \gb}{}^{\gc}  X^i_{\gc} \; ,
\nonumber
\end{equation}
where $C_{\ga \gb}{}^{\gc} $ denote the structure constants
of the Bianchi type under consideration. The basis dual
to $X^i_{\ga}$, defined with
$ X^i_{\ga} \; \; \chi_i^{\gb} = \delta_{\ga}^{\gb}$, satisfies,
\begin{equation}
2 \partial_{[i} \chi_{j]}^{\ga} = - C_{\gb \gc}{}^{\ga}
 \chi_i^{\gb} \chi_j^{\gc} \; .
\nonumber
\end{equation}
Without loss of generality, one may set
\begin{equation}
C_{\ga \gb}{}^{\gc} = \epsilon_{\ga \gb \gd} M^{\gd \gc} +
2 \delta_{[\ga}^{\gc} V_{\gb ]} \; ,
\nonumber
\end{equation}
with $M^{\ga \gb}$ symmetric.
 From the Jacobi identities,
it follows that $M^{\ga \gb} V_{\gb} = 0 $. The Bianchi classification
is given by the vanishing or not of $V_{\ga}$,
and the rank and the signature of $M^{\ga \gb}$, subject to this condition.
Models with $V_{\ga} = 0$, $V_{\ga} \ne 0$, are called type A,
and type B, respectively. In the type A models,
the most popular are Bianchi I, selected by $M^{\ga \gb} = 0$,
and Bianchi IX,
selected by  $ M^{\ga \gb} = \delta^{\ga \gb}$.

\vspace{.5cm}

\noindent{\bf 4. Super-Minisuperspace}

\vspace{.3cm}

The Jacobson phase space variables for N=1 supergravity
may be expanded
with respect to the kinematical triad $X^i_{\ga}$ (or $\chi_i^{\gb}$), as,
\begin{eqnarray}
A_{i A}{}^B &=& A_{\ga A}{}^B \chi_i^{\ga}, \nonumber\\
\st^{iAB} &=& (det\chi )\; \s^{\ga AB} X^i_{\ga}, \nonumber \\
\p_i^A  &=& \p_{\ga}^A \chi_i^{\ga},  \nonumber \\
\pt^{iA} &=& (det\chi ) \pi^{\ga A} X^i_{\ga}, \nonumber
\end{eqnarray}
where $ det \chi $ denotes the determinant of $\chi_a^{\ga}$, which is
introduced in order to de-densitize $\st^{iAB}$ and $\pt^{iA}$.
Note that $\p_{\ga}^A$ and $\pi^{\gb B}$ are anti-commuting.
All of the spatial dependence of the fields is contained in the
kinematical quantities $X^i_{\ga}$ and $\chi_i^{\gb}$.
The phase space has been reduced from 30 degrees of freedom per space
point to  only 30 global degrees of freedom.

The Poisson brackets take the form
\begin{eqnarray}
\{ \s^{\ga AB} , A_{\gb MN} \} &=&  {i \over \sqrt{2}} \delta_{\gb}^{\ga} \;
\delta_{(M}{}^A
\delta_{N)}{}^B , \label{eq:pbh1}\\
\{ \pi^{\ga A} , \p_{\gb M}  \} &=&   {i \over \sqrt{2}} \delta_{\gb }^{\ga }
\; \delta_M{}^A .
\label{eq:pbh2}
\end{eqnarray}

Inserting the expansion in the supersymmetry constraints,
they reduce to
\begin{eqnarray}
S^A &:=&  ( C_{\gb \ga}{}^{\gb} \epsilon^{AB}
+ A_{\ga}{}^{AB} ) \pi^{\ga}_B = 0,
\label{eq:susyb1} \\
S^{\dagger A} &:=&  ( - {1 \over 2}\;
C_{\ga \gb}{}^{\gc} \delta_C^D
+  A_{\ga C}{}^D \delta_{\gb}^{\gc} ) \psi_{\gc D}
\s^{[\ga | AB |} \s^{\gb]}{}_B{}^C = 0.
\label{eq:susyb2}
\end{eqnarray}
The diffeomorphism and $SL(2,C)$ constraints become
\begin{eqnarray}
H^{AB} &:=& -  ( \s^{\ga } \s^{\gb} C_{\ga \gb}{}^{\gc} A_{\gc } )^{ BA} +
(\s^{[\ga } \s^{\gb ]} A_{\ga } A_{\gb } )^{ BA}  \nonumber \\
&-& (\pi^{\ga }  \s^{\gb }  C_{\ga \gb}{}^{\gc} \p_{\gc } )\e^{AB}
+ 2 (\pi^{[\ga }\s^{\gb ]} A_{\ga } \p_{\gb } )\e^{AB}  \nonumber \\
&-& (\pi^{\ga } C_{\ga \gb}{}^{\gc} \p_{\gc }) \s^{\gb AB}
+ 2 (\pi^{[\ga } A_{\ga} \p_{\gb}) \s^{\gb ]AB} = 0,
\label{eq:hdiffeo}\\
J^{AB} :&=&   C_{\gb \ga}{}^{\gb} \s^{\ga AB}
+ 2 A_{\ga}{}^{C(A} \s^{| \ga | B)}{}_C
- \pi^{\ga (A} \psi_{\ga}^{B)} = 0 .
\label{eq:hgauss}
\end{eqnarray}
Note that we have rescaled the constraints by the appropriate
power of $det \chi$ to de-densitize them.

In general, it is not guaranteed that the truncation to
the homogenous solutions will preserve the Poisson algebra
of the constraints.  For our purposes, it is sufficient
to consider the algebra of the supersymmetry generators.
It is automatic
that $ \{ S^A , S^B \} = 0 $. A short calculation shows that
also the right-handed generators continue to close into themselves,
\ie that $\{ S^{\dagger A} , S^{\dagger B } \}
= C^{AB}{}_C (\s \p) S^{\dagger C }$. On the other hand,
the mixed bracket yields
\begin{eqnarray}
\{  S^A , S^{\dagger B } \} &=& {i \over 2 \sqrt{2}} H^{AB} \nonumber \\
&+& {i \over 2 \sqrt{2}} C_{\gb \ga}{}^{\gb} C_{\gc \gd}{}^{\ga} (\s^{\gc}
\s^{\gd})^{BA}
+  {i \over \sqrt{2}} C_{\gb \gc}{}^{\gb} ( \s^{[\gc } \s^{\gd ]}
A_{\gd})^{BA}.
\end{eqnarray}
The extra terms on the right hand side vanish when
$C_{\gb \ga}{}^{\gb} = 0$, \ie for type A Bianchi models.

The supersymmetry constraints fail to close for  Bianchi
type B. For this reason, in addition to the  (possibly related) well known
difficulties one
encounters
in giving a hamiltonian formulation of these models,
in the following  we will restrict ourselves to Bianchi
models of type A, \ie  we will
assume that   $C_{\gb \ga}{}^{\gb} = 0$.

For this case, the  supersymmetry constraints reduce further to
\begin{eqnarray}
S^{\dagger A} &=&  - {1 \over 2} \e_{\ga \gb \gd} M^{\gd \gc} (\s^{\ga}
\s^{\gb} \psi_{\gc } )^A
+ ( \s^{[\ga } \s^{\gb]} A_{\ga} \psi_{\gb} )^A, \label{eq:rsusy1}\\
S^A &=& A_{\ga}{}^{AB} \pi^{\ga}_B .   \label{eq:rsusy2}
\end{eqnarray}

At this point, it is tempting to specialize further, \eg by considering
$\s^{\ga AB} , A_{\ga AB}$ diagonal, when considered as three by three matrices
\cite{DEH,DEIX}.
We will resist this temptation, since it
complicates, rather than simplifies, the form of the supersymmetry constraints.

\vspace{1cm}

\noindent{\bf 5. Quantization. Triad Representation}

\vspace{.5cm}

In this section, we turn to the quantization of type
A Bianchi models.
In the triad representation, quantum states may be
represented by wavefunctions that
depend on the triad, and on the gravitino field,
$\Psi = \Psi (\s, \psi)$. The Poisson bracket (\ref{eq:pbh1})
turns into commutators of operators, while (\ref{eq:pbh2}) into
anti-commutators. The variables $\s^{\ga AB}$ and $\psi{\ga }^A$ are
considered as `position' operators.  We use the standard notation
of denoting operators with a hat. Their momenta
may be represented  with
\begin{eqnarray}
\hat{A}_{\ga}{}^{AB} \Psi &=&  {1  \over \sqrt{2}}  {\delta \Psi
\over \delta \s^{\ga}{}_{AB} },  \nonumber \\
\hat{\pi}^{\ga}{}_A \Psi &=&  {1  \over \sqrt{2}}
{\delta \Psi \over \delta \psi_{\ga}{}^A }. \nonumber
\end{eqnarray}

In the translation of the supersymmetry constraints to their quantum version,
the
issue of factor ordering arises in the second term of (\ref{eq:rsusy1}).
Schematically, three possibilities are available:
(i) $\s \s A$; (ii) $\s A \s$; (iii) $A \s \s$. To accomodate this
ambiguity, we write the quantum version of
(\ref{eq:rsusy1}) in the form
\begin{equation}
\hat{S}^{\dagger A} \Psi (\s , \psi ) =
[ - {1 \over 2} \e_{\ga \gb \gd} M^{\gd \gc} (\s^{\ga}
\s^{\gb} \psi_{\gc })^A
+ {1  \over \sqrt{2}}  (\s^{[\ga} \s^{\gb]} {\delta \over \delta \s^{\ga} }
\psi_{\gb})^A
+ {1  \over \sqrt{2}} \alpha \s^{\ga AC} \psi_{\ga C} ] \Psi
= 0
\label{eq:qc3}
\end{equation}
where the constant $\alpha $ characterizes the factor ordering,  \eg
$\alpha = 0,1,2$ for the orderings (i), (ii), (iii), respectively.

The left-handed generator (\ref{eq:rsusy2}) takes the form,
\begin{equation}
\hat{S}^A \Psi (\s , \psi) =   {1 \over 2} {\delta^2 \Psi
\over \delta \s^{\ga}{}_{AB} \delta \psi_{\ga}{}^B }
= 0 , \label{eq:qc1}
\end{equation}

In addition,  a physical state must also satisfy
\begin{equation}
\hat{J}^{AB} \Psi = 0,
\label{eq:qsl}
\end{equation}
 \ie be invariant under
$SL(2,C)$ rotations. It follows
that the wavefunction must be an $SL(2,C)$ scalar.
(In $\hat{J}^{AB}$, we assume an appropriate factor ordering  so that it
generates
$SL(2,C)$ rotations of the wavefunction.)

As mentioned in the introduction,
since the hamiltonian system is finite dimensional,
 the wavefunction may be expanded in even powers of the
gravitino fields, symbolically as follows,
\begin{equation}
\Psi (\s ,\psi ) =  \Psi_{(0)} (\s ) + \Psi_{(2)} (\s ,\psi ) +
\Psi_{(4)} (\s ,\psi ) + \Psi_{(6)} (\s ,\psi ),
\label{eq:wf}
\end{equation}
where the subscript indicates the number of gravitino fields.
Only even powers appear because of $SL(2,C)$ invariance.
Since the super-symmetry generators do not mix fermionic
number, this decomposition permits us to solve
the quantum constraints (\ref{eq:qc3},\ref{eq:qc1},\ref{eq:qsl})
order by order.

\vspace{.5cm}

\noindent{\bf 5.1 Bosonic states}

\vspace{.3cm}

The first term in the wavefunction (\ref{eq:wf}) is $\Psi_{(0)} (\s)$. From
$SL(2,C)$ invariance, it can depend on $\s^{\ga}{}_{AB}$ only
in the combination $ h^{\ga \gb} := tr (\s^{\ga} \s^{\gb}) $,
\ie $\Psi_{(0)} =  \Psi_{(0)} (h)$.

In this representation,   $\Psi_{(0)} $ satisfies (\ref{eq:qc1}) automatically.
We are left with  (\ref{eq:qc3}). The gravitino field
is arbitrary,  so we
can peel it off. Multiplying through by $\s^{\ge}{}_{AD}$ gives,
after some index reshuffling,
\begin{equation}
[ - {1 \over 2} \epsilon_{\ga \gb \gc} M^{\gc \gd}
Tr (\s^{\ga} \s^{\gb} \s^{\ge } )
+ {1  \over \sqrt{2}} Tr ( \s^e \s^{[\ga} \s^{\gd]} {\delta \over \delta
\s^{\ga}})
+ {1  \over \sqrt{2}} \alpha h^{\gd \ge}] \Psi_{(0)} = 0.
\label{eq:int1}
\end{equation}
Using the identities
\begin{eqnarray}
\sqrt{2} Tr (\s^{\ga} \s^{\gb} \s^{\gc } ) &=& -  h^{-1/2}
\e^{\ga \gb \gc}, \nonumber \\
2 \s^{[\ga|AB|} \s^{\gd]}{}_{B(C} \s^{\ge}_{D)A} &=&
 h^{\ge \ga} \s^{\gd}{}_{CD}
- h^{\ge \gd} \s^{\ga}{}_{CD} , \nonumber
\end{eqnarray}
where we denote with $h$ the determinant of $h_{\ga \gb }$,
and the chain rule, $Tr (\s^{\ga} \delta / \delta \s^{\gb} ) =
2 h^{\ga \gc} \delta / \delta h^{\gb \gc} $,  equation (\ref{eq:int1}) can be
put in the form,
\begin{equation}
[  (h^{\ga \gb} h^{\gc \gd} - h^{\ga \gc} h^{\gb \gd} ) {\delta \over
\delta h^{\gc \gd}} + \alpha h^{\ga \gb} +  h^{-1/2}
M^{\ga \gb} ] \Psi_{(0)} = 0 .
\end{equation}
This equation can be easily integrated to give
\begin{equation}
\Psi_{(0)} (h) =  k_{(0)} \; h^{\alpha/2} \; \; exp \{ 2 i F \},
\end{equation}
where $k_{(0)}$ is a constant, and
\begin{equation}
F :=
-{ i \over 2} h^{-1/2} h_{\ga \gb}
M^{\ga \gb}  .
\label{eq:gen}
\end{equation}

The exponent of the prefactor $h^{\alpha/2}$ can be adjusted by changing the
factor ordering. Until an inner product is available, we cannot
appeal to physical arguments to select a preferred factor ordering.

The quantity $F$ is the bosonic part of the homogenous specialization
of the
generating functional of the canonical transformation
from the triad extended ADM phase space to Jacobson's
\cite{GK} .

It is interesting to observe that we could have chosen as well the `momentum'
 representation for the fermionic fields. In this representation,
for a bosonic wavefunction, $\hat{S}^{\dagger A} \Psi_{(0)} (\s) = 0$ is
satisfied
automatically. $\hat{S}^A \Psi_{(0)} = 0$,
implies $ \delta \Psi_{(0)}/ \delta \s^{\ga AB} = 0$, \ie that $\Psi_{(0)}$ is
constant.
This should serve as a reminder that physical states can take
different forms, depending crucially on the representation
chosen. In the following, for concreteness, and to remain close
to previous treatments, we will stick to the `position' representation.

\vspace{.5cm}

\noindent{\bf 5.2 Fermionic states}

\vspace{.3cm}

We turn now to the solution of
(\ref{eq:qc3},\ref{eq:qc1},\ref{eq:qsl}) for wavefunctions that
depend on the gravitino field.

Our first step is to identify the irreducible
spin 1/2 and spin 3/2 parts of the gravitino field.
Let,
\begin{equation}
\psi^{ABC}  := \psi_{\ga}^A \s^{\ga BC} = \psi^{A(BC)}.
\end{equation}
Then,
\begin{equation}
\psi^{ABC} = \rho^{ABC} + \epsilon^{A(B} \beta^{C)}
\end{equation}
where $\rho^{ABC} = \rho^{(ABC)}$, represents the spin 3/2 part, and $\beta^A =
(2/3) \psi_B{}^{AB}$, the spin 1/2 part.
In what follows it will be a key fact that $\rho^{ABC}$ and $\beta^A$
can be specified independently.

The wavefunction can depend on the gravitino field
only in Lorentz invariant combinations. The
only non-vanishing possibilities are
\begin{eqnarray}
\mbox{second order} &:&  [\rho^2 ] := \rho^{ABC} \rho_{ABC}, \; \; \;
[\beta^2] : = \beta^A \beta_A , \nonumber\\
\mbox{fourth order} &:& [\rho^2]^2 , \; \; \; [\rho^2] [\beta^2],  \nonumber
 \\
\mbox{sixth order} &:&  [\rho^2]^2 [\beta^2]. \nonumber
\end{eqnarray}

Note that in terms of the phase space variables one has,
\begin{displaymath}
[\rho^2 ] = { 2 \over 3 } [ \psi_{\ga}{}^A \psi_{\gb A} h^{\ga \gb} -
(2h)^{-1/2}  \psi_{\ga}{}^A \psi_{\gb}{}^B
\epsilon^{\ga \gb \gc} \s_{\gc AB}  \},
\end{displaymath}
\begin{displaymath}
[ \beta^2 ]
= { 2 \over 9 } \{ \psi_{\ga}{}^A \psi_{\gb A} h^{\ga \gb} +
\sqrt{2}
h^{-1/2}  \psi_{\ga}{}^A \psi_{\gb}{}^B
\epsilon^{\ga \gb \gc } \s _{\gc AB}  \}.
\end{displaymath}

Then, the most general expression for wavefunctions
that depend on the gravitino field is of the form
\begin{eqnarray}
\Psi_{(2)} (\s , \psi ) &=&  F_1 (h) [\rho^2 ] + F_2 (h) [\beta^2], \\
\Psi_{(4)} (\s , \psi ) &=& G_1 (h) [\rho^2]^2  + G_2 (h) [\rho^2] [\beta^2],
\\
\Psi_{(6)}(\s , \psi ) &=&  H (h) [\rho^2]^2 [\beta^2],
\end{eqnarray}
where the functions $F,G,H$ depend on $\s^{\ga AB}$ only in the combination
$h^{\ga \gb } = Tr ( \s^{\ga} \s^{\gb} )$.

It is convenient at this point to express (\ref{eq:qc3}) in terms of
the independent quantities $\beta^A, \rho^{ABC}$,
\begin{eqnarray}
 \sqrt{2} \hat{S}^{\dagger A} \Psi = &-& h^{-1/2} M^{\ga \gb} \s_{\ga}{}^{AB}
\s_{\gb}^{CD} \rho_{BCD} \Psi
+  \s^{\ga(A}{}_B \rho^{C)DB} {\delta \Psi \over \delta \s^{\ga CD}}
\nonumber \\
&+& {1 \over 2} \beta^A [ - h^{-1/2} M^{\ga \gb} h_{\ga \gb}
+ 3   \alpha + \s^{\ga CD}
{\delta  \over \delta \s^{\ga CD}} ] \Psi = 0.
\label{eq:qci}
\end{eqnarray}
It turns out that, in general,
the only physical states of second and fourth order
are the trivial ones, \ie
\begin{equation}
\Psi_{(2)} = 0 = \Psi_{(4)}.
\end{equation}
The reason is that these states must satisfy both quantum
supersymmetry equations. When one isolates all the independent terms in these
equations,
they turn out to be too many relations to be satisfied at the same time.
The exceptions are given by special cases like Bianchi I, and a
specific factor ordering.

For concreteness, we show explicitly how it goes for $\Psi_{(2)}$.
The computation for $\Psi_{(4)}$ follows the same pattern, and comes
to the same conclusion.

Consider first ,
\begin{equation}
\hat{S}^A \Psi_{(2)} = 2
 \s^{\ga AB} \s^{\gb CD} \rho_{BCD}
{\delta F_1 \over \delta h^{\ga \gb } }
- { 2\over 3}\beta^A
[ h^{\ga \gb} {\delta F_2 \over h^{\ga \gb }} + 4 F_2 - 3 F_1 ] \}
= 0
\end{equation}
Since $\rho^{ABC}$ and $\beta^A$ can be specified independently, their
coefficients must vanish separately.
The vanishing of the $\rho^{ABC}$ coefficient implies
\begin{equation}
{\delta F_1 \over \delta h^{\ga \gb } } \propto h_{\ga \gb},
\end{equation}
while the vanishing of the $\beta$ coefficient gives
\begin{equation}
   h^{\ga \gb} {\delta F_2 \over \delta h^{\ga \gb }} + 4 F_2 - 3 F_1 = 0.
\end{equation}
It is not necessary to integrate these equations. Using them, (\ref{eq:qc3})
takes the form,
\begin{eqnarray}
\sqrt{2} \hat{S}^{\dagger A} \Psi_{(2)}
&=& -  [\rho^2]  h^{-1/2} M^{\ga \gb} \s_{\ga}{}^{AB}
\s_{\gb}^{CD} \rho_{BCD} F_1 \nonumber \\
&-&  {1 \over 2} \; [\rho^2] \beta^A \{ [ h^{-1/2} M^{\ga \gb} h_{\ga
\gb} + 3  \alpha + 2   ] F_1
+ {1 \over 6} F_2
+ h^{\ga \gb} {\delta F_1 \over \delta h^{\ga \gb} } \}
\nonumber \\
&-& [\beta^2] \{  h^{-1/2} M^{\ga \gb} \s_{\ga}{}^{AB}
\s_{\gb}^{CD} \rho_{BCD} F_2
+ \s^{\ga AB} \s^{\gb CD} \rho_{BCD}
{\delta F_2 \over \delta h^{\ga \gb } } \} = 0 . \nonumber
\end{eqnarray}
Each line of this equation must vanish separately. If
$M^{\ga \gb} \ne 0$, it follows immediately that $F_1 = F_2 = 0$,
\ie that $\Psi_{(2)} = 0 $. For $M^{\ga \gb} = 0$, \ie Bianchi I,
there is a factor ordering  such that
$F_1 = F_2 = k_{(2)} h^{1/3}$, with $k_{(2)}$ a constant.

We come now to the last term in the wavefunction (\ref{eq:wf}),
\ie $\Psi_{(6)}$. Note that
 $\hat{S}^{\dagger A} \Psi_{(6)} = 0$ is
 identically satisfied, since it is of seventh order in six anti-commuting
 quantities. We are left with
 \begin{eqnarray}
 \hat{S}^A \Psi_{(6)} = &+& 4
 [\rho^2 ] [\beta^2 ] \s^{\ga AB} \s^{\gb CD} \rho_{BCD} {\delta H
\over \delta h^{\ga \gb }} \nonumber \\
&-&  2  \beta^A [\rho^2]^2
 \{ {1 \over 3} h^{\ga \gb }   {\delta H \over \delta h^{\ga \gb}}
+ 3 H \}  = 0 .
 \end{eqnarray}
 Each line must vanish separately.
 Integrating  gives
 \begin{equation}
 H = k_{(6)} h^{1/3},
 \end{equation}
 where $k_{(6)}$ is a constant.
 Note that this term in the wavefunction is
 independent both of the specific Bianchi type A, and of the factor
ordering.

 \vspace{.5cm}

\noindent{\bf ACKNOWLEDGEMENTS}

\vspace{.3cm}
We thank T. Jacobson for reading the manuscript and for many
valuable suggestions.
It is a pleasure to thank O. Obreg\'on  and M. Ryan for discussions.
We also  thank O. Obreg\'on for bringing Ref. \cite{ATY} to our attention.

\bibliographystyle{plain}

\end{document}